\documentclass[preprint,aps,floats,showpacs]{revtex4}
\usepackage{graphicx}
\def\upper{\hbox{${}^{\perp}$}\!}
\def\uppar{\hbox{${}^{\parallel}$}\!}

\begin{document}

\title{Pre-big bang on the brane}

\author{Stefano Foffa}
\affiliation{
D\'epartment de Physique Th\'eorique, Universit\'e de Gen\`eve,
\\24, quai Ernest-Ansermet, CH-1211 Gen\`eve 4, Switzerland
\\email: foffa@amorgos.unige.ch}

\begin{abstract}
The equations of motion and junction conditions for a gravi-dilaton brane
world scenario are studied in the string frame.
It is shown that they allow Kasner-like solutions on the brane,
which makes the dynamics of the brane very similar to the low
curvature phase of pre-big bang cosmology.
Analogies and differences of this scenario with the Randall-Sundrum
one and with the standard bulk pre-big bang dynamics are also discussed.
\end{abstract}
\pacs{98.80.Cq, 04.50.+h, 11.25.Mj}

\maketitle

\section{Introduction}
The pre-big bang model \cite{prebig} is an attempt at constructing
a nonsingular cosmological scenario based on string theory.
A decade of work \cite{decade} has produced a model
which contains inflation, is nonsingular, and has a number of
distinctive phenomenological predictions.

This model has some drawbacks, too. On the phenomenological side, it predicts
a generally blue spectrum of cosmological perturbations \cite{pert};
this feature, which is very welcome for what concerns the possible detection
of gravitational waves of cosmological origin \cite{mag},
determines on the other hand a huge suppression of scalar
perturbations at large scale. Some modifications of the minimal version of
the model have been recently proposed in order to render its
predictions for the scalar perturbation spectrum compatible with the
large scale inhomogeneities observed in the cosmic microwave background
radiation (CMBR) \cite{scalar}.

On the theoretical side, a very serious obstacle is
related to the study of the high curvature, strongly coupled phase, and
to the dilaton stabilization: henceforth I will refer
to all these theoretical challenges as the graceful exit problem.
Though some physical mechanisms (such as the inclusion of $\alpha'$ corrections
and loop terms, or entropy considerations) for resolving this problem
have been found and studied, also with encouraging results, it is still true
that a full comprehension of the ``stringy'' phase of the model has not
yet been established, due mainly to the poor technical control one has
on this phase.

Since no significant progress has been made in the last few years for solving
the graceful exit problem, it is unlikely that the techniques used so far
will help further. Rather, a new physical mechanism, or a new framework,
or at least a change of perspective could be quite helpful at this point.
This is exactly the spirit of this paper, in which the pre-big bang
dynamics will be reproduced in a brane-world scenario.
The general idea (which goes beyond the purpose of the present work) is that
the previously mentioned problems could assume a different form in a brane
world context, and maybe become less difficult if treated
in terms of brane world-related techniques (such as AdS-CFT).
In this sense, the approach of this paper could be considered similar
to the one taken in the so-called ekpyrotic model \cite{ekpy}.

Started as an alternative to compactification \cite{RaSu},
brane world models have been extensively studied recently:
new explicit solutions were discovered and the original Randall-Sundrum
picture was extended to more general situations like
the inclusion of matter and of a cosmological constant on the brane
(see \cite{brane} and references therein), and the relation of this
scenario with the AdS-CFT conjecture has been further investigated
\cite{Gub}.
This scenario being inspired by string theory, a further natural extension
has been the inclusion of a scalar field, both in the
Einstein \cite{brdil} and in the string \cite{strfr} frame.

In the present paper I will work exclusively in the string frame:
since the goal is to find out whether, and under which conditions,
a pre-big bang phase can take place on the brane, I find that a string frame
treatment is the most appropriate one.

The action will be displayed and the bulk and boundary equations will be 
written. Then, following a standard technique, the bulk equations will be
projected on the brane and studied.
It will turn out that, despite the presence of the dilaton (which induces
energy nonconservation on the brane), it is possible in some situations to
study the brane dynamics in a rather self-contained way, without needing to
know the full bulk solution.

Then an homogeneous and isotropic ansatz will be made on the brane, and the 
relative equations of motion will be written. They are found to be very
similar to ordinary bulk vacuum gravi-dilaton equations,
and indeed they admit similar solutions, which are written, displayed and
commented.
Some final remarks conclude this work.

\section{The equations in the string frame}
Let us consider a $D$ dimensional spacetime, in which one dimension has been
compactified on a $S^{1}/Z_{2}$ orbifold (it is assumed that the other
$10$-$D$ spatial dimensions are small, compact and frozen).
This orbifold has the topology of a segment,
and the $Z_{2}$ symmetric points are effectively the boundaries
of spacetime; I will assume that one brane is located at one of these points.
This is the simplest and most common choice in brane world scenarios, though
more general configurations, involving more complicated junction conditions,
can also be considered (see \cite{RutFil} for more details).
The action governing the dynamics of such a system is
\begin{eqnarray}\label{action}
{\cal S}=\int {\rm d}^D x \sqrt{-g}{\rm e}^{-\phi}\left[{\cal R} +
\left({\cal D}\phi^2\right) + {\cal L}_{bulk}\right]-
\int{\rm d}^{D-1} y \sqrt{-h}{\rm e}^{-\phi}
\left[2 K^{\pm}+ {\cal L}_{brane}\right]\, .
\end{eqnarray}
Throughout the paper I will denote with ${\cal R}$ and ${\cal D}_{\mu}$,
respectively, the $D$-dimensional Ricci tensor and the covariant
derivative with respect to the bulk metric $g_{\mu\nu}$, while the symbols
$R$ and $\nabla_{\mu}$ will indicate the
Ricci tensor of the brane and the covariant derivative with respect
to the induced metric $h_{\mu\nu}$.
Moreover, the second fundamental form $K_{\mu\nu}$ is defined as
$K_{\mu\nu}=h^{\rho}_{\mu}h^{\sigma}_{\nu}{\cal D}_{\rho}n_{\sigma}$,
with $n_{\mu}$ a unitary vector field pointing {\it into} the bulk,
orthogonal to the boundary at the boundary location, and defined in the bulk
by the condition $n^{\rho}{\cal D}_{\rho}n_{\mu}=0$
(this definition is always valid at least in the neighborhoods of
the boundary); with these definitions,
one has $g_{\mu\nu}=h_{\mu\nu}+n_{\mu}n_{\nu}$.

Having clarified the notations, one can read in (\ref{action}) that the
bulk part is the usual low energy effective action for gravity and dilaton
in the string frame plus a generic bulk matter contribution denoted as
${\cal L}_{bulk}$, while the boundary, as well as the Gibbons-Hawking term
$K^{\pm}$ (in which the $\pm$ means that one must take the sum of the
value of that term at both sides of the brane), contains also the
brane-matter lagrangian ${\cal L}_{brane}$.

Variation of (\ref{action}) with respect to the bulk metric and to the dilaton
gives directly the bulk equations of motion
\begin{eqnarray}\label{eqbulk1}
& &{\cal R}_{\mu\nu} + {\cal D}_{\mu}{\cal D}_{\nu}\phi=
T_{\mu\nu} - \frac{1}{2}g_{\mu\nu} T^{\phi}\, ,\\
\label{eqbulk2}
& &{\cal R} - \left({\cal D}\phi\right)^2 + 2 {\cal D}^2 \phi= - T^{\phi}\, ,
\end{eqnarray}
and the boundary junction conditions
(here specialized to the $Z_2$ symmetric case)
\begin{eqnarray}\label{junct}
2K_{\mu\nu}=K_{\mu\nu}^{\pm}=\tau_{\mu\nu}-\frac{\tau^{\phi}}{2}h_{\mu\nu}\, ;
\quad 2 \upper{\cal D}\phi\equiv\left(n\cdot{\cal D}\phi\right)^{\pm}=
\tau-\frac{D-2}{2}\tau^{\phi}\, .
\end{eqnarray}
In the previous equations $T_{\mu\nu}=-\frac{1}{\sqrt{-g}}
\frac{\delta\sqrt{-g}{\cal L}_{bulk}}{\delta g^{\mu\nu}}$,
$T^{\phi}=-{\rm e}^{\phi}\frac{\delta{\rm e}^{-\phi}{\cal L}_{bulk}}
{\delta \phi}$, and analogous definitions hold
for $\tau_{\mu\nu}$ and $\tau^{\phi}$, with ${\cal L}_{bulk}$ and $\sqrt{-g} $
replaced by ${\cal L}_{brane}$ and  $\sqrt{-h}$, respectively. 
The junction conditions look quite different from the usual
Einstein frame ones for a gravi-scalar system; this is due the to the fact
that in the string frame the equations for $K_{\mu\nu}$ and $\phi$
are mixed together. A derivation of the junction conditions (\ref{junct})
in the string frame is presented in the Appendix.

It is convenient to split the quantities appearing in the bulk equations
(\ref{eqbulk1},\ref{eqbulk2}) into their components parallel
and orthogonal to the brane.
Thus, given a symmetric tensor $S_{\mu\nu}$, I define
$\uppar S_{\mu\nu}\equiv h^{\rho}_{\mu}h^{\sigma}_{\nu}
S_{\rho\sigma}$ and $\upper S\equiv n^{\rho}n^{\sigma}S_{\rho\sigma}$,
so that the trace of $S$ can be written as
$S_{\mu\nu}g^{\mu\nu}=\upper S + \uppar S$.
Using these definitions, the equations of motion can be projected
and rewritten as follows:
\begin{eqnarray}\label{eqpar}
& &\uppar{\cal R}_{\mu\nu}+ \uppar
{\cal D}_{\mu}{\cal D}_{\nu}\phi= \uppar T_{\mu\nu}- \frac{1}{2}T^{\phi}
h_{\mu\nu}\, ,\\
\label{eqper}
& &\upper{\cal R} + \upper{\cal D}^2 \phi = \upper T- \frac{1}{2}T^{\phi}\, ,\\
\label{eqdil}
& & \uppar {\cal R}+ \upper {\cal R} - \left(\upper{\cal D}\phi\right)^2
- h^{\mu\nu}{\cal D}_{\mu}\phi{\cal D}_{\nu}\phi + 2\upper{\cal D}^2 \phi
+ 2 \uppar{\cal D}^2 \phi=-T^{\phi}
\end{eqnarray}
(there is also a mixed $\perp\parallel$ component which has not been reported
here, since it will not be needed in the following).
This form is convenient for studying the system from the point of view of
the brane. Indeed, one can express the $(D-1)$-dimensional Ricci tensor
and the $(D-1)$-dimensional double covariant derivative of $\phi$
on the brane in terms of the various projections of the corresponding
$D$-dimensional quantities and of the projection of the Weyl tensor,
$E_{\mu\nu}={\cal C}_{\alpha\beta\gamma\delta}n^{\alpha}n^{\gamma}
h^{\beta}_{\mu}h^{\delta}_{\nu}$:
\begin{eqnarray}\label{GauCod}
&&R_{\mu\nu}=\frac{D-3}{D-2}\uppar{\cal R}_{\mu\nu}
-\frac{h_{\mu\nu}}{(D-2)(D-1)}\left[(D-2)\upper{\cal R}
-\ \uppar{\cal R}\right] + K K_{\mu\nu} - K_{\mu\alpha}
K^{\alpha}_{\nu} - E_{\mu\nu}\nonumber\\
&&\quad\quad\quad\quad\quad\quad\quad
\quad\quad\quad\quad\quad\quad\quad
\quad\quad\quad\quad\quad\quad\quad
\quad\quad\quad\quad\quad
{\it [Gauss\!-\!Codacci]}\, ,\\
\label{boxDd}
&&\nabla_{\mu}\nabla_{\nu}\phi=\uppar{\cal D}_{\mu}{\cal D}_{\nu}\phi
- K_{\mu\nu} \upper{\cal D} \phi \, .
\end{eqnarray}

Now, from (\ref{GauCod}) one can write the brane Einstein tensor in terms of
bulk curvature tensors and of $K_{\mu\nu}$, then one uses the projected
equations of motion to express the bulk curvature tensors in terms of dilaton
derivatives and of bulk sources, and at last decomposes the dilaton derivatives
into their brane and bulk part thanks to eq.~(\ref{boxDd}).
Finally, after having moved on the l.~h.~s.~all the brane tensors,
one is left with
\begin{eqnarray}\label{00comp}
R_{\mu\nu}&-&\frac{h_{\mu\nu}}{2}\left[R+2\frac{D-3}{D-2}\nabla^2\phi
-\frac{D-3}{D-1}\left(\nabla\phi\right)^2\right]+
\frac{D-3}{D-2}\nabla_{\mu}\nabla_{\nu}\phi=\nonumber \\
&=&\frac{D-3}{D-2}\left(\uppar T_{\mu\nu} -
\frac{h_{\mu\nu}}{D-1}\uppar T\right)+\frac{D-3}{D-1}h_{\mu\nu}\upper T
- \frac{D-3}{2(D-1)}h_{\mu\nu}\left(\upper{\cal D}\phi\right)^2 +
\nonumber\\
&+&\left(h^{\rho}_{\mu}h^{\sigma}_{\nu} - \frac{1}{2}h^{\rho\sigma}
h_{\mu\nu}\right)\left(KK_{\rho\sigma}-K_{\rho}^{\alpha}K_{\sigma\alpha}\right)
-\frac{D-3}{D-2}\left(K_{\mu\nu} -h_{\mu\nu}K\right)\upper{\cal D}\phi
- E_{\mu\nu}.
\end{eqnarray}
If the dilaton derivatives are neglected in the previous expression,
one is formally left with the usual brane world Einstein equations
(see, for example, the second reference of \cite{brane} for a comparison),
in which the brane curvature tensors are expressed in terms of the bulk and
brane sources (the latter enter through the extrinsic curvature tensor
because of the junction conditions) and of $E_{\mu\nu}$, which has the
peculiarity of not being determined by the bulk equations of motion.
In general the form of the projected Weyl tensor can be inferred from the
symmetries of the system, and thus the brane world Einstein equations can
be studied in a rather self-contained way.
The addition of the dilaton does not change (so far) this state of things,
since also the terms containing $\upper {\cal D}\phi$ can be expressed in
terms of brane sources via the junction conditions (\ref{junct}).

It is interesting to take the trace of this equation, which gives
\begin{eqnarray}\label{bradil}
R-\left(\nabla\phi\right)^2 + 2\nabla^2 \phi= - 2 \upper T +
K^2 - K_{\mu\nu}K^{\mu\nu} - 2 K \upper{\cal D}\phi +
\left(\upper{\cal D}\phi\right)^2\, .
\end{eqnarray}
The structure tensor of the l.~h.~s.~of this equation is formally the same
as the usual bulk dilaton equation of motion
in the string frame (\ref{eqbulk2}); this fact will make our purpose
of looking for pre-big bang-like solutions on the brane easier.

Since eqns.~(\ref{00comp},\ref{bradil}) are not independent, I need another
scalar equation of motion; this can be found, for example, by taking the
trace of (\ref{eqpar}) and then using eqns.~(\ref{GauCod},\ref{boxDd}),
as well as the other bulk equations of motion, in order to finally reach
the following result:
\begin{eqnarray}\label{trace}
R&+&\nabla^2\phi=\uppar T -\upper T -\frac{D-2}{2}T^{\phi} + K^2 -
K_{\alpha\beta}K^{\alpha\beta} - K \upper {\cal D}\phi +
\upper{\cal D}^2 \phi\, .
\end{eqnarray}
As before, the structure tensor on the l.~h.~s.~is the same of
the bulk gravi-dilaton equation (\ref{eqbulk1}).
What makes this equation qualitatively different from (\ref{bradil})
is the presence of the term $\upper{\cal D}^2 \phi$.
Contrarily to all the other terms on the r.~h.~s.,
this is not expressible in terms of sources because it is not determined
by the junction conditions. Rather, in order to know the value of
$\upper{\cal D}^2 \phi$ at the brane location, one should in
principle solve the full set of bulk equations.
Thus, differently from what happens in the brane Einstein equations,
for a gravi-dilaton system generally it is not possible to study the brane
evolution in a self-contained way, without knowing the bulk dynamics.
This fact has already been pointed out in the first works on
gravi-scalar brane models \cite{brdil} and it is related to the fact that
in such systems the dilaton dynamics generally induces energy-momentum
nonconservation on the brane.

However it will be shown in the next section that in some specific
situations the term  $\upper{\cal D}^2 \phi$ can be determined {\it at the
brane} without solving the bulk equations, thus allowing a selfcontained
description of the brane evolution.

Before proceeding further, it is perhaps worth pointing out that,
despite the fact that there is some freedom in choosing the form of the
evolution equations for our brane world model (and indeed this freedom was
used to write the equations in a way that will result the most appropriate
for the  purposes of this work), this does not mean that there is any
arbitrariness in them.
That is, there are not other independent ways to write down combinations
of the brane curvature tensors and of the derivative of the dilaton
on the brane in terms of the bulk and brane sources, of
$E_{\mu\nu}$ and of $\upper {\cal D}^2 \phi$.

\section{Looking for pre-big bang solutions}

\subsection{Specification of the model}
Let us now choose the matter content of the model. 
For the bulk part I take ${\cal L}_{bulk}=-2\Lambda$, so that
the matter term is given simply by a cosmological constant:
\begin{eqnarray}\label{Tmunu}
T_{\mu\nu}=-\Lambda g_{\mu\nu}\, ;\quad\quad T^{\phi}=-2\Lambda\, .
\end{eqnarray}
As to the brane matter part, the following term is considered:
\begin{eqnarray}\label{brten}
{\cal L}_{brane}=2\lambda {\rm e}^{\xi\Phi+\phi}\, ,
\end{eqnarray}
where $\Phi$ is the $10-$dimensional dilaton, which can be expressed
in terms of the $D-$dimensional one $\phi$ by means of the relation
${\rm e}^{-\phi}={\rm e}^{-\Phi}{\cal V}$, ${\cal V}$
being the volume (in string units) of the compact ($10$-$D$)-dimensional space,
which is supposed to be a number of order $1$.
The parametrization has been chosen in such a way that for
$\xi=-1/2$ one recovers the usual D-brane tension behavior
$\sim {\rm e}^{-\Phi/2}\sim1/g_{10}$, while for $\xi=-1$ the brane tension
term has the same dilaton prefactor ${\rm e}^{-\phi}$ as all
the other terms in the action, thus making this case the most appropriate
one for a comparison with the standard Randall-Sundrum scenario.

Given eq.~(\ref{brten}), the brane matter tensors are
\begin{eqnarray}\label{taumunu}
\tau_{\mu\nu}=\lambda h_{\mu\nu}{\rm e}^{(\xi+1)\phi} {\cal V}^{\xi}\, ,
\quad\quad \tau^{\phi}=-2\lambda\xi{\rm e}^{(\xi+1)\phi} {\cal V}^{\xi}\, ,
\end{eqnarray}
and the junction conditions (\ref{junct}) read:
\begin{eqnarray}\label{KDp}
2 K_{\mu\nu}=(\xi+1)\lambda h_{\mu\nu}{\rm e}^{(\xi+1)\phi} {\cal V}^{\xi}\, ,
\quad\quad 2 \upper{\cal D}\phi=[(D-2)(\xi+1)+1]\lambda
{\rm e}^{(\xi+1)\phi} {\cal V}^{\xi}\, .
\end{eqnarray}
Moreover, I have in mind to study the most symmetric configuration, in which
the brane metric and the dilaton are homogeneous on the brane.
In this case only three equations are needed. The $00$ component of
(\ref{00comp}), which does not contain second time derivatives,
gives a constraint on the initial conditions, as usually happens in
general relativity.
The remaining two equations, which contain second time derivatives
of the dilaton and of the scale factor on the brane,
are given by eq.~(\ref{trace}) and by any of
the $ij$ components of (\ref{00comp}) or, better, by its trace,
eq.~(\ref{bradil}).

One can now rewrite the full set of equations,
after having expressed everything in terms 
of $\Lambda$ and $\lambda$:
\begin{eqnarray}\label{fin00}
R_{\mu\nu}&-&\frac{h_{\mu\nu}}{2}\left[R + 2 \frac{D-3}{D-2}\nabla^2\phi
-\frac{D-3}{D-1} \left(\nabla\phi\right)^2\right]+
\frac{D-3}{D-2}\nabla_{\mu}\nabla_{\nu}\phi=\nonumber \\
&=&-\frac{(D-3)}{D-1}\left[\Lambda + \frac{\lambda^2}{8}F_{\xi}(D)
{\cal V}^{2\xi}{\rm e}^{2(\xi+1)\phi}\right]
h_{\mu\nu}- E_{\mu\nu}\quad {\rm (\mbox{00}\ component)}\, ,\\
\label{findil}
R&+&2\nabla^2\phi -\left(\nabla\phi\right)^2=2\Lambda + \frac{\lambda^2}{4}
F_{\xi}(D){\cal V}^{2\xi}{\rm e}^{2(\xi+1)\phi}\, ,\\
\label{fintrace}
R&+&\nabla^2\phi=\upper{\cal D}^2 \phi-\frac{\lambda^2}{4}(D-1)
\left(\xi+1\right){\cal V}^{2\xi}{\rm e}^{2(\xi+1)\phi}\, ,
\end{eqnarray}
with $F_{\xi}(D)=\xi^2 - (D-1) (\xi + 1)^2$.

For $\xi>-1$ (which includes also the D-brane case, $\xi=-1/2$),
the brane matter terms are multiplied by a positive power of the
$D$-dimensional coupling constant $g_{D}={\rm e}^{\phi/2}$; thus they
should be considered strong coupling corrections, and can be disregarded
in the small coupling limit.
On the contrary, for $\xi=-1$, these terms are as relevant as the others
and should be taken into account. The case $\xi<-1$ will not be considered
here.

After having noticed that each of the different bulk sources appears
in only one equation (the projection of the
Weyl tensor only in the constraint equation, and the term
$\upper{\cal D}^2 \phi$ only in the last one), one observes also
that the same combination of $\Lambda$ and $\lambda$ appears
in the first two equations.

One can take advantage of this fact by tuning the parameters of the
model in such a way that this particular combination vanishes.
This requirement is analogous to the condition of vanishing
cosmological constant on the brane, which is usually imposed in pure gravity
brane world models.

For $\xi=-1$ one has $F_{\xi}(D)=1$ and thus the requirement is
$\lambda^2=-8 {\cal V}^2\Lambda$.
It should be noted that, even for ${\cal V}=1$, this condition
is slightly different from the one found in the Randall-Sundrum model,
which was $\lambda^2=-\frac{8(D-2)}{D-1}\Lambda$.
This could be surprising in light of the previous comment about the
fact that eq.~(\ref{00comp}) reduces to the standard brane world case if
the dilaton is neglected, a fact which does not seem to happen in
eq.~(\ref{fin00}), nor for $\xi=-1$.
The explanation is that the latter (and its trace, eq.~(\ref{findil}))
has been obtained after making use of the junction conditions
(\ref{junct}).
As a matter of fact, one of the differences between the present system and 
the pure gravity one is that here the dilaton couples to the brane;
because of this, the formal limit of neglecting the dilaton derivatives
cannot be done after one has already used the junction conditions.
In other words, here a different condition is obtained because the
physical system is different. 

Coming to the case $\xi>-1$, all the $\lambda$ terms should be neglected in
the small coupling limit, and thus the vanishing brane cosmological
constant requirement reduces to $\Lambda=0$. Under such a condition, there is
not confinement of a ($D$-$1$)-dimensional graviton on the brane,
and this should certainly be considered a problem if one wants to recover
Einstein gravity on the brane. However here the goal is to find
a pre-big bang solution, which does not need to satisfy such a constraint,
and so the requirement $\Lambda=0$ (which is, in my opinion, slightly
less awkward than the RS one) is not phenomenologically problematic.

Coming back to the equations, it has just been shown that one can tune 
the parameters in order to make the r.~h.~s.~of eq.~(\ref{findil}) vanish.
Now let us push things further by considering the particular case $E_{00}=0$:
in this case the only source term we are left with is the
$\upper{\cal D}^2 \phi$ term that appears in eq.~(\ref{fintrace})
(the $\lambda$ term in that equation is absent even for $\xi=-1$ because
its coefficient vanishes).

To summarize, after these assumptions on the sources
one remains with a system of equations which has many similarities
with the usual {\it vacuum} bulk gravi-dilaton system, the only differences
being: (i) the different tensor structure of the constraint equation, and
(ii) the presence of a source term in eq.~(\ref{fintrace}). It will now be
shown that these two ``problems'' are actually one the ``solution'' of the
other.

\subsection{Homogeneous solution}
It is now time to look for an explicit solution of this system.
The following ansatz is made for the induced metric and for the dilaton
on the brane:
\begin{eqnarray}\label{ansatz}
{\rm d}s^2_{brane}=-{\rm d}t^2 + a^2(t){\rm d}\vec{x}^2\, ,\quad
\phi_{brane}=\varphi(t)\, ,
\end{eqnarray}
and these expressions are substituted into eqns.~(\ref{fin00}, \ref{findil},
\ref{fintrace}) thus giving, after some rearrangements in the two dynamical
equations:
\begin{eqnarray}\label{00ans}
&&(D-2)(D-1)H^2 + \dot{\varphi}^2 - 2 (D-1)\dot{\varphi}H=0\, ,\\
\label{phiddot}
&&\ddot{\varphi}=\dot{\varphi}^2- (D-2)\dot{\varphi}H +
\upper{\cal D}^2 \phi \, ,\\
\label{Hdot}
&&2(D-2)\dot{H}=\dot{\varphi}^2-(D-2)(D-1)H^2 + 2 \upper{\cal D}^2 \phi\, ,
\end{eqnarray}
with $H\equiv\dot{a}/a$ and $\dot{\varphi}\equiv{\rm d}\varphi/{\rm d}t$.

\begin{figure}
\begin{center}
\includegraphics[width=2.6in, angle=-90]{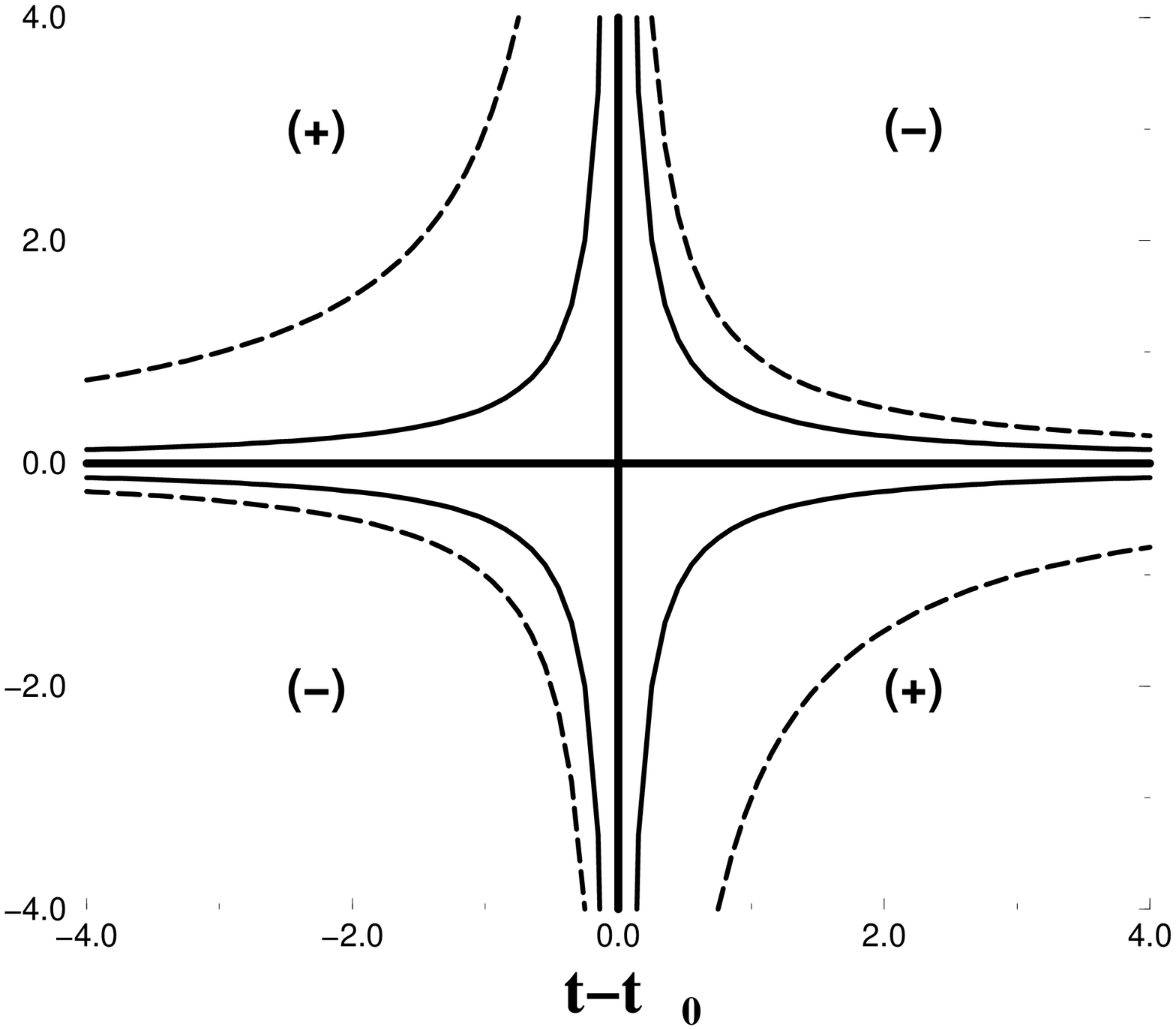}
\includegraphics[width=2.6in, angle=-90]{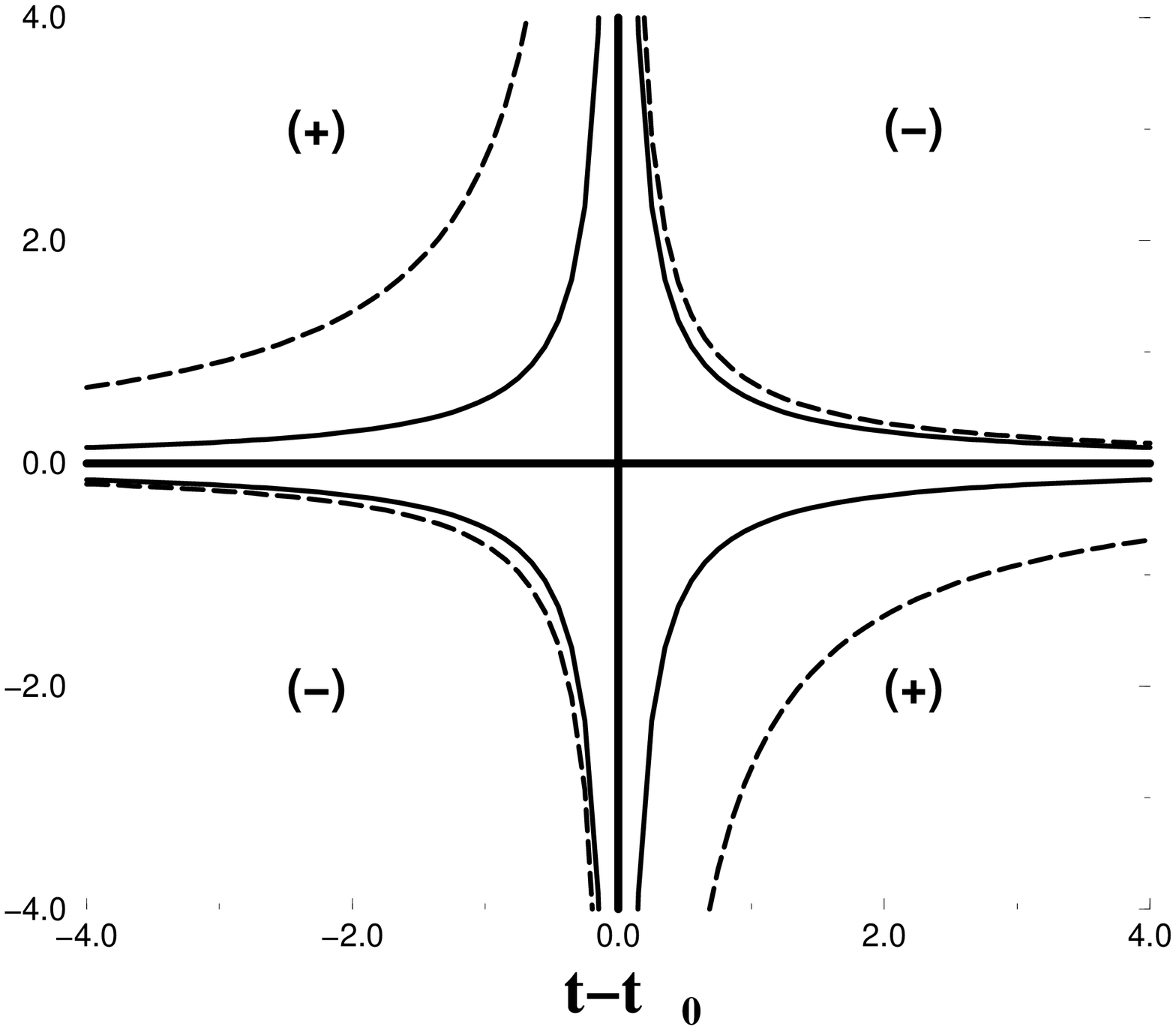}
{\bf (a)}\hspace{12cm}{\bf (b)}
\caption{\label{figura} On the left (a) the solutions (\ref{easy}) for $H$
(solid line) and $\dot{\varphi}$ (dashed); both the $(+)$ and $(-)$
branches are displayed. On the right (b) the analogous pre- and post-big
bang solutions for the bulk vacuum equations, in the same number of spatial
dimensions. The data displayed in these graphs correspond to the case $d=3$;
for $d>3$ the solutions are qualitatively the same.}
\end{center}
\end{figure}

A crucial observation should be made at this point:
despite the fact that we cannot know
{\it a\ priori} what is the behavior of $\phi$ in the bulk,
we know that the constraint (\ref{00ans}) must be satisfied
on the brane.
This information can be used to express the value at the brane of
$\upper{\cal D}^2\phi$ in terms of $\dot{\varphi}$ and $H$ only.
This is done by taking the time derivative of the constraint
equation and by using the dynamical equations to express $\ddot{\varphi}$,
$\dot{H}$ in terms of $\dot{\varphi}$, $H$ and $\upper{\cal D}^2\phi$,
thus obtaining the following condition:
\begin{eqnarray}\label{cond}
2\dot{\varphi}\upper{\cal D}^2 \phi=(D-3)
\dot{\varphi}^3 - [(D-2)(D-1)]^2 H^3 + (D-2)(3D-5) H \dot{\varphi}
[(D-1)H -\dot{\varphi}]\, .
\end{eqnarray}
This condition can now be used to re-express the dynamical equations in terms
of $H$ and $\varphi$ only. As it happens in the ordinary bulk vacuum case,
the resulting equations are easily solved by means of
the following ansatz, which corresponds to a Kasner-like solution
\begin{eqnarray}\label{easy}
\left\{
\begin{array}{l}
\dot{H}=\beta^{\pm} H^2 \\
\dot{\varphi}=\alpha^{\pm} H \\
\end{array}
\right.\quad
\Rightarrow\quad
\left\{\begin{array}{l}
H=-\frac{1}{\beta^{\pm} (t-t_0)}\\
\dot{\varphi}=-\frac{\alpha^{\pm}}{\beta^{\pm}}\frac{1}{t-t_0}\ \ \, .
\end{array}
\right.
\end{eqnarray}
The two $D$-dependent couples of coefficients $(\alpha^{\pm},\beta^{\pm})$
correspond to two different branches (conventionally indicated as $(\pm)$
according to the sign of $\beta$) of the solution, just like in the bulk
vacuum string cosmology equations. In the latter case it was $\alpha^{\pm}=
d\pm\sqrt{d}$, $\beta^\pm=\pm\sqrt{d}$, with $d$ the number of spatial
dimensions (for a comparison with our case, one should take $d=D-2$).
In terms of $d$, the coefficients for the brane world solution (\ref{easy})
turn out to be only slightly different from the bulk case:
\begin{eqnarray}\label{alpbet}
\alpha^{\pm}=d+1\pm \sqrt{d+1}\, ,\quad\quad
\beta^{\pm}=\pm\sqrt{d+1}\, . 
\end{eqnarray}
The solutions (\ref{easy}) are plotted in fig.~\ref{figura}.
As can be seen, the bulk and brane world behaviors are pratically the same:
in both cases $H$ and $\dot{\varphi}$ are always increasing [decreasing]
in the $(+)$ [$(-)$] branch, and they are positive [negative] for
$t<t_0$ and viceversa.
If one chooses to look at the $(+)$ branch for $t<t_0$
the initial state is flat and weakly coupled, and evolves towards a
singularity going through a period of inflationary expansion and increasing
coupling: this is the pre-big bang phase.
If one considers the $(-)$ branch for $t>t_0$, one sees a period of
decelerated expansion which originates from a singularity, just like in
standard cosmology.

At $t$ approaches $t_0$, the solution is not reliable anymore because
it goes beyond the validity regime of the low energy effective action
(\ref{action}): at large curvature the role of $\alpha'$ corrections
should be taken into account, while as $\varphi$ increases the dynamics
is certainly modified by perturbative (loops) and nonperturbative strong
coupling effects among which, in the case $\xi>-1$, one should also
include the terms proportional to ${\rm e}^{(\xi+1)\phi}$ that appear in
eqns.~(\ref{fin00}, \ref{findil}, \ref{fintrace}).
The graceful exit problem in pre-big bang cosmology consists of connecting
the two zeroth order branches of the solution into a whole nonsingular
expanding evolution.
As mentioned in the introduction, this problem has not yet found
a satisfactory solution.

The fact that the vacuum pre-big bang dynamics have been reproduced in a
brane world model shows that the graceful exit problem can be
rephrased in a brane world context.
For instance, the inclusion of $\alpha'$ corrections is expected to damp
the singularity also in the brane world case (some indications have already
been found in \cite{higher}). In the same spirit, in the present context
the eventual presence of a singularity in the bulk at a finite distance from
the brane should be regarded as irrelevant, since at high curvature
the solution found from the low energy action is not valid anymore.

Finally, for what concerns the dilaton stabilization, its realization in this
context corresponds to the brane moving through a region of constant
dilaton. Of course this issue should also be studied in detail in order
to understand whether or not there are specific technical obstacles
to the realization of this scenario.

\section{Conclusions}
I have just shown that a pre-big bang like dynamics can take place, under
rather mild assumptions, also in a brane world model.
The equations for such a model have been written in the
string frame, and their tensor structure reminds that of
the well known equations for a bulk gravi-dilaton system.
For a particular choice of the matter content of the model
the similarity becomes even more evident and, if the bulk Weyl tensor
is set to zero, the system has the same behavior of the vacuum bulk
pre-big bang cosmology.

Further investigation of the analogies and differences between the bulk and
brane realization of pre-big bang, with particular reference to the
high curvature and strong coupling phase and to the graceful exit problem,
are left for future work.

\begin{center}
{\bf ACKNOWLEDGMENTS}
\end{center}
I wish to thank Ruth Durrer for some interesting remarks
on the manuscript, and Michele Maggiore for useful comments
and suggestions.

\begin{center}
{\bf APPENDIX}
\end{center}
Here the junction conditions for the string frame action (\ref{action})
are derived. The derivation presented here is a modification of the
Einstein frame one given in \cite{jc}, to which the reader is addressed
for further details.

The variation of the action is the following:
\begin{eqnarray}\label{var1}
\delta{\cal S}&=&\int {\rm d}^D x {\rm e}^{-\phi}\left[\sqrt{-g}\left(
\left\{{\cal R}_{\mu\nu}- \frac{1}{2}g_{\mu\nu}\left[{\cal R} -
\left({\cal D}\phi\right)^2 + 2 {\cal D}^2 \phi\right] +
{\cal D}_{\mu}{\cal D}_{\nu}\phi- T_{\mu\nu}\right\}\delta g^{\mu\nu}+\right.
\right.\nonumber\\
&-&\left.\left\{{\cal R} - \left({\cal D}\phi\right)^2 +
2 {\cal D}^2 \phi + T^{\phi} \right\}\delta\phi\right)+\nonumber\\
&+&\left.\left({\rm e}^{-\phi}\sqrt{-g}
\left\{g^{\mu\nu}\delta\Gamma^{\rho}_{\mu\nu}-g^{\mu\rho}\delta
\Gamma^{\nu}_{\mu\nu}+
\left[g^{\nu\rho}{\cal D}^{\mu}\phi-g^{\mu\nu}{\cal D}^{\rho}\phi\right]
\delta g_{\mu\nu} + \left[2{\cal D}^{\rho}\phi\right]\delta\phi
\right\}\right)_{\, ,\rho}\right]+\nonumber\\
&-&\int {\rm d}^{D-1} y \sqrt{-h}{\rm e}^{-\phi}\left[2 \delta K^{\pm}
+K^{\pm} h^{\mu\nu} \delta g_{\mu\nu}- \tau_{\mu\nu} \delta g^{\mu\nu}
- \left(2 K^{\pm}+ \tau^{\phi}\right)\delta\phi\right]\, .
\end{eqnarray}
The first two lines give the bulk equations of motion.
As to the third line, it can be tranformed into an integral on the boundary
by making use of the Gauss theorem which,
with our definition of $n_{\mu}$, can be written as
\begin{eqnarray}\label{gauss}
\int {\rm d}^{D} x \left(\sqrt{-g} V^{\rho}\right)_{\, ,\rho}\longrightarrow
- \int {\rm d}^{D-1} y \sqrt{-h} \left(n_{\rho}V^{\rho}\right)^{\pm}\, .
\end{eqnarray}
After some rearrangements in order to express the $\delta \Gamma$'s in terms
of derivatives of $\delta g$, one finds that the third line becomes
\begin{eqnarray}\label{third}
-\int {\rm d}^{D-1} y \sqrt{-h}{\rm e}^{-\phi}\left[
n^\rho h^{\mu\nu}\left({\cal D}_{\mu}\delta g_{\rho\nu}-
{\cal D}_{\rho} \delta g_{\mu\nu}\right) 
+\left(n^{\mu}{\cal D}^{\nu}\phi-g^{\mu\nu}n_{\rho}{\cal D}^{\rho}\phi\right)
\delta g_{\mu\nu} + 2 n_{\rho}{\cal D}^{\rho}\phi \delta \phi
\right]^{\pm}\, .
\end{eqnarray}
Coming to the fourth line of equation (\ref{var1}), it can be computed
by knowing the definition of $K_{\mu\nu}$ in terms of $n_{\mu}$, and by
recalling that the normalization condition for $n_{\mu}$ implies
\begin{eqnarray}\label{dn}
\delta n_{\mu}=\frac{1}{2}n_{\mu}n^{\rho}n^{\sigma}\delta g_{\rho\sigma}\, ,
\end{eqnarray}
thus giving
\begin{eqnarray}\label{dK}
\delta K=K_{\mu\nu}\delta g^{\mu\nu} +
\frac{1}{2}K n^{\rho}n^{\sigma}\delta g_{\rho\sigma}
-h^{\mu\nu}n^{\rho}\left({\cal D}_{\mu}\delta g_{\nu\rho}-
\frac{1}{2} {\cal D}_{\rho}\delta g_{\mu\nu}\right)\, .
\end{eqnarray}
Now everything can be put together and the boundary part of $\delta{\cal S}$
becomes:
\begin{eqnarray}\label{dsb}
\delta {\cal S}_{b}&=&\int{\rm d}^{D-1} y \sqrt{-h}{\rm e}^{-\phi}
\left[\left(h^{\mu\nu}n^{\rho}{\cal D}_{\mu}\delta g_{\nu\rho}\right)^{\pm}+
\left(n_{\mu}{\cal D}_{\nu}\phi-g_{\mu\nu}n_{\rho}{\cal D}^{\rho}\phi
+K g_{\mu\nu} - 2K_{\mu\nu}\right)^{\pm}\delta g^{\mu\nu}\right.\nonumber\\
&+&\left.\tau_{\mu\nu}\delta g^{\mu\nu} +
2 \left(K- n_{\rho}{\cal D}^{\rho}\phi\right)^{\pm}\delta\phi
+ \tau^{\phi}\delta \phi\right]\, .
\end{eqnarray}
The first term can be rewritten in the following way:
\begin{eqnarray}\label{part}
\sqrt{-h}\left[{\cal D}_{\mu}\left({\rm e}^{-\phi}h^{\mu\nu}n^{\rho}
\delta g_{\nu\rho}\right)-\delta g_{\nu\rho}
{\cal D}_{\mu}\left({\rm e}^{-\phi}h^{\mu\nu}n^{\rho}\right)\right]^{\pm}\, ,
\end{eqnarray}
and the first part vanishes because ${\cal D}_{\mu}V^{\mu}\equiv
\nabla_{\mu}V^{\mu} + n^{\mu}n^{\nu}{\cal D}_{\mu}V_{\nu}$
when $V_{\mu}$ is tangential to the brane, and because the boundary
of the boundary is zero.

After having expanded the derivative of the second term of (\ref{part}),
one is finally left with
\begin{eqnarray}\label{final}
\delta {\cal S}_{b}&=&\int{\rm d}^{D-1} y \sqrt{-h}{\rm e}^{-\phi}
\left\{\left[K^{\pm} h_{\mu\nu} - K_{\mu\nu}^{\pm} -h_{\mu\nu}
\left(n_{\rho}{\cal D}^{\rho}\phi\right)^{\pm}
+\tau_{\mu\nu}\right] \delta g^{\mu\nu} +\right.\nonumber\\
&+&\left. \left[2 \left(K- n_{\rho}{\cal D}^{\rho}\phi\right)^{\pm}+ 
\tau^{\phi}\right]\delta \phi \right\}\, ,
\end{eqnarray}
from which the junction conditions (\ref{junct}) can be
straightforwardly derived.
\end{document}